\documentclass[hidelinks,conference]{IEEEtran}
\usepackage[T1]{fontenc}
\usepackage[latin9]{inputenc}
\usepackage{verbatim}
\usepackage{amsmath}
\usepackage{graphicx}

\usepackage{xcolor}
\usepackage[colorlinks=true, linkcolor=darkcyan, citecolor=black, urlcolor=black, pdfborder={0 0 0}]{hyperref}    

\makeatletter
\usepackage{amsfonts}\usepackage{algorithmic}
\usepackage{array}
\usepackage{textcomp}
\usepackage{stfloats}
\usepackage{url}
\usepackage{verbatim}
\usepackage{soul}\usepackage{threeparttable}
\usepackage{algorithmic}
\def\BibTeX{{\rm B\kern-.05em{\sc i\kern-.025em b}\kern-.08em
    T\kern-.1667em\lower.7ex\hbox{E}\kern-.125emX}}
\usepackage{balance}
\usepackage{cite}
\definecolor{darkgreen}{rgb}{0.0, 0.673, 0.51}
\definecolor{darkcyan}{rgb}{0.0, 0.35, 0.55}
\makeatother

\usepackage{fancyhdr}

\fancypagestyle{firstpage}{%
  \fancyhf{} 
  
  \fancyhead[L]{\scriptsize \copyright~ 2024 IEEE. Personal use of this material is permitted.  Permission from IEEE must be obtained for all other uses, in any current or future media, including reprinting/republishing this material for advertising or promotional purposes, creating new collective works, for resale or redistribution to servers or lists, or reuse of any copyrighted component of this work in other works. 
  DOI: 10.1109/ICCT62411.2024.10946352}%
}

\begin{document}
\title{Digital Linearizer Based on 1-Bit Quantizations}
\author{\IEEEauthorblockN{Deijany Rodriguez Linares}
	\IEEEauthorblockA{Link\"oping University\\
		Department of Electrical Engineering\\
		581  83 Link\"oping, Sweden\\
		Email: deijany.rodriguez.linares@liu.se}\\
	
	\and
	\IEEEauthorblockN{H\aa kan Johansson}
	\IEEEauthorblockA{Link\"oping University\\
		Department of Electrical Engineering\\
		581  83 Link\"oping, Sweden\\
		Email: hakan.johansson@liu.se}\\
}
\maketitle
\thispagestyle{firstpage} 

\begin{abstract}
This paper introduces a novel low-complexity memoryless linearizer
for suppression of distortion in analog frontends. It is based on
our recently introduced linearizer which is inspired by neural networks,
but with orders-of-magnitude lower complexity than conventional neural-networks
considered in this context, and it can also outperform the conventional
parallel memoryless Hammerstein linearizer. Further, it can be designed
through matrix inversion and thereby the costly and time consuming
numerical optimization traditionally used when training neural networks
is avoided. The linearizer proposed in this paper is different in
that it uses 1-bit quantizations as nonlinear activation functions
and different bias values. These features enable a look-up table implementation
which eliminates all but one of the multiplications and additions
required for the linearization. Extensive simulations and comparisons
are included in the paper, for distorted multi-tone signals and bandpass
filtered white noise, which demonstrate the efficacy of the proposed
linearizer.

\end{abstract}

\section{Introduction\label{sec:Intro}}

Digital linearization (post-correction and pre-distortion) are used for
suppressing nonlinearities (distortion) emanating from imperfections
in analog circuits and systems. In this paper, the focus is on digital
linearization needed in e.g. analog front-ends on the receiver
side in communication systems. The nonlinearities emanate from the
analog-to-digital converters (ADCs) and its preceding components including
filters, amplifiers, and mixers. In addition to suppressing the nonlinearities,
it is important to develop low-complexity digital linearizers to enable
energy-efficient hardware implementations, especially in systems with
very high data rates. Efficient digital linearizers also enable the
use of ADCs with relaxed linearity requirements and resolutions (fewer
bits) which can reduce their energy consumption substantially \cite{Murmann_2021a}. 

This paper concerns frequency-independent distortion for which one
can use memoryless polynomial modeling and linearization\footnote{Memoryless nonlinearity modeling and linearization is typically sufficient
for narrow to medium analog bandwidths and resolutions. To reach higher
resolutions over wider frequency bands, one may need to incorporate
memory (subfilters) in the modeling and linearization.}. Recently, \cite{deiro23} introduced a low-complexity memoryless
linearizer (Fig. \ref{Flo:proposed-scheme} in Section \ref{sec:Proposed-linearizer})
that is inspired by neural networks, but has orders-of-magnitude lower
complexity than conventional neural-network schemes considered in
this context \cite{DENG_202063,Peng_2021}, and it can be designed
through matrix inversion, thereby avoiding the costly and time-consuming
numerical optimization that is traditionally used when training neural
networks. It was also demonstrated in \cite{deiro23} that it can outperform the conventional
parallel memoryless Hammerstein linearizer \cite{Chen_95} (Fig. \ref{Flo:Hammerstein-scheme}
in Section \ref{sec:Signal-model-and-proposed-linearizer}). Nevertheless,
the linearizer introduced in \cite{deiro23} still requires several
multiplications and additions per linearized sample. 

In this paper, a new linearizer is proposed which enables a look-up
table implementation, which eliminates all but one of the multiplications
and additions required for the linearization. The proposed linearizer
is based on the one in \cite{deiro23} but differs in two ways. Firstly,
the nonlinear activation functions are here 1-bit quantizers instead
of the rectified linear unit (ReLU) or modulus operations adopted
in \cite{deiro23}. Secondly, instead of determining the bias values
in the design as in \cite{deiro23}, they are here a-priori selected
carefully. An additional advantage of the proposed linearizer, over
that in \cite{deiro23}, is that it requires only one design (one
matrix inversion). In \cite{deiro23}, several designs are used to
find the optimal bias values.

Following this introduction, Section \ref{sec:Signal-model-and-proposed-linearizer}
briefly recapitulates the linearization problem and reviews the Hammerstein
linearizer and the one in \cite{deiro23}. Section \ref{sec:Proposed-linearizer}
considers the proposed linearizer and its design. Section \ref{sec:Evaluation-and-Results}
provides evaluations and comparisons, demonstrating the efficacy of
the proposed linearizer. Finally, Section \ref{sec:Conclusions} concludes
the paper.

\section{Distortion and Linearization \label{sec:Signal-model-and-proposed-linearizer}}
Consider a desired discrete-time signal $x(n)=x_{a}(nT)$, representing
a sampled version of an analog signal $x_{a}(t)$ with a uniform sampling
interval $T$. In practice, the output of an ADC will not be $x(n)$
but a distorted version of it, say $v(n)$. In this paper, it is assumed
that the distortion is memoryless, in which case $v(n)$ is commonly
modeled as a memoryless polynomial according to \cite{Chen_95}
\begin{equation}
v(n)=a_{0}+a_{1}x(n)+\sum_{p=2}^{P}a_{p}x^{p}(n),\label{eq:signal-model}
\end{equation}
where $a_{0}$ is a constant (offset), $a_{1}$ is a linear-distortion
coefficient and $a_{p}$, $p=2,3,\ldots,P$, are nonlinear-distortion
coefficients. Additionally, the signal contains quantization noise,
but it is here excluded from the mathematical expressions for the
sake of simplicity. Before proceeding, it is also stressed that the
proposed linearizer (Section \ref{sec:Proposed-linearizer}) as well
as the existing ones to be reviewed below do not require that the
distorted signal is in the form of \eqref{eq:signal-model}. In the
simulations in this paper, the model in \eqref{eq:signal-model} is
used for generating sets of test and evaluation signals in order to
assess the performance of the different linearizers.

\begin{figure}
\centering \includegraphics[scale=0.55]{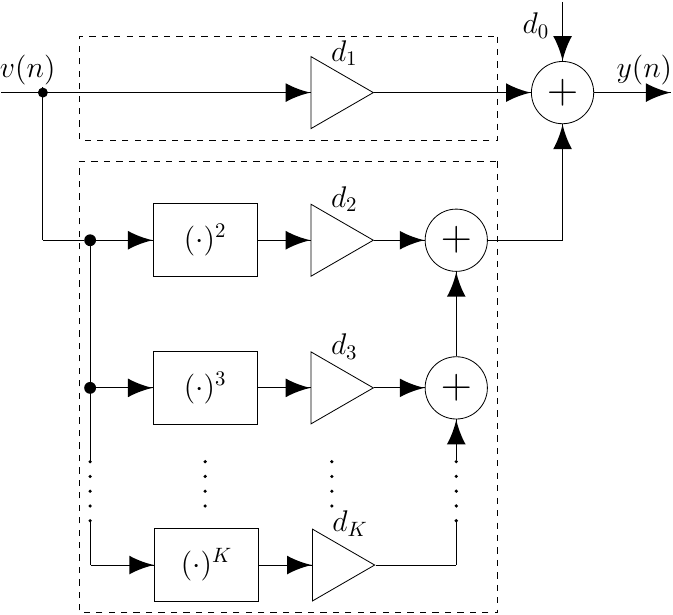}
\caption{Hammerstein linearizer with the upper (lower) dashed box indicating
the linear branch (nonlinear branches).}
\label{Flo:Hammerstein-scheme}
\end{figure}

Given the distorted signal $v(n)$, the linearization amounts to generating
a compensated signal, say $y(n)$, in which the distortion has been
suppressed (ideally removed). In the traditional Hammerstein linearizer,
illustrated in Fig. \ref{Flo:Hammerstein-scheme}, $y(n)$ is generated
as
\begin{equation}
y(n)=d_{0}+d_{1}v(n)+\sum_{k=2}^{K}d_{k}v^{k}(n).
\end{equation}
In total, this linearizer requires $2K-1$ multiplications and $K$
additions per corrected output sample, with $K$ multiplications to
produce $d_{k}v^{k}(n)$ and $K-1$ multiplications for all the $v^{k}(n)$
computations. Alternatively, one may use the so-called Wiener linearizer
\cite{Chen_95} where the multiplications by $d_{k}$, $k=2,3,\ldots,K$,
seen in Fig. \ref{Flo:Hammerstein-scheme}, are carried out before
the operations $v^{k}(n)$, but it does not change the implementation
complexity and performance.

\begin{figure}
\begin{centering}
\includegraphics[scale=0.55]{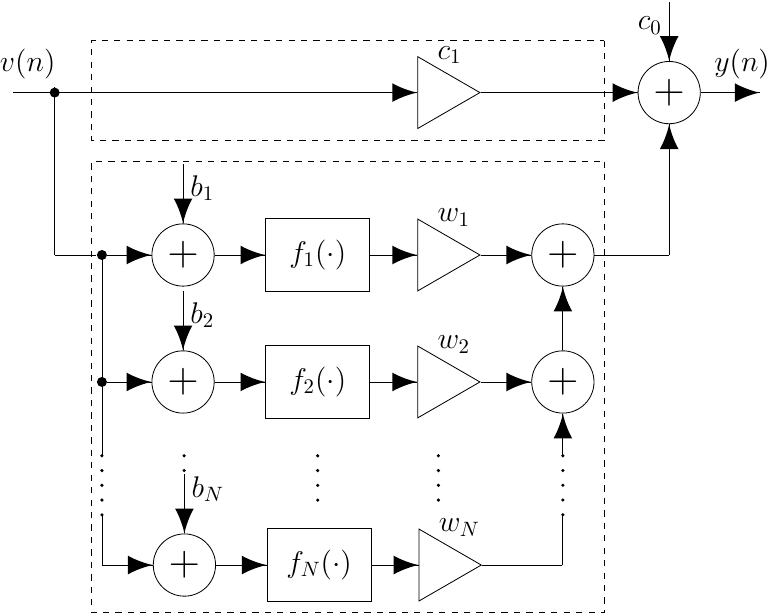}
\par\end{centering}
\caption{Linearizer in \cite{deiro23} with the upper (lower) dashed box indicating
the linear branch (nonlinear branches).}
\label{Flo:proposed-scheme}
\end{figure}

Recently, the linearizer seen in Fig. \ref{Flo:proposed-scheme} was
introduced \cite{deiro23}. The compensated signal $y(n)$ is then
generated as
\begin{equation}
y(n)=c_{0}+c_{1}v(n)+\sum_{m=1}^{N}w_{m}f_{m}(v(n)+b_{m})\label{eq:prop_linearizer}
\end{equation}
with $f_{m}$, $m=1,2,\ldots,N$, representing a set of $N$ nonlinear
functions. In \cite{deiro23}, the ReLU operations
$f_{m}(v(n))=\max\{0,v(n)+b_{m}\}$ or modulus operations $f_{m}(v(n)+b_{m})=|v(n)+b_{m}|$
were used due to their simplicity in hardware implementations \cite{Tarver_2019}.
Further, the bias values $b_{m}$ were uniformly distributed between
$-b_{\max}$ and $b_{\max}$, i.e., $b_{m}=-b_{\max}+2(m-1)b_{\max}/(N-1)$,
where the value of $b_{\max}$ was determined in the design. It was
demonstrated in \cite{deiro23}, that the linearizer in Fig. \ref{Flo:proposed-scheme}
can outperform the Hammerstein linearizer even when the nonlinearities
have been generated through the memoryless polynomial model in \eqref{eq:signal-model}.
However, the linearizer in Fig. \ref{Flo:proposed-scheme} still requires
$N+1$ multiplications and $2N+1$ additions per corrected output
sample.

\section{Proposed Linearizer \label{sec:Proposed-linearizer}}

In this paper, a new linearizer is proposed which enables a look-up
table implementation, thereby eliminating the multiplications and
additions in the $N$-branch nonlinearity part of the linearizer in
Fig. \ref{Flo:proposed-scheme}. It is based
on the one in Fig. \ref{Flo:proposed-scheme} but it has two distinct
differences. The first one is that the nonlinear functions are here
1-bit quantizations according to
\begin{equation}
f_{m}\big(v(n)+b_{m}\big)=\begin{cases}
1, & v(n)+b_{m}\geq0,\\
0, & v(n)+b_{m}<0.
\end{cases}\label{eq:one-bit-quantizer}
\end{equation}
The second difference is that, instead of determining $b_{\max}$
in the design \cite{deiro23}, it is here selected as
\begin{equation}
b_{\max}=\frac{N-1}{N+1}.\label{eq:b_max}
\end{equation}
 The bias values $b_{m}$ are then again uniformly distributed between
$-b_{\max}$ and $b_{\max}$. Utilizing \eqref{eq:b_max}, $b_{m}$
are thus chosen as
\begin{equation}
b_{m}=-1+\frac{2m}{N+1},\,m=1,2,\ldots,N.\label{eq:b_m}
\end{equation}

The implication of the choices in \eqref{eq:one-bit-quantizer}--\eqref{eq:b_m} is two-fold. Firstly, for each sample index $n$,
since the output of each nonlinear operation $f_{m}\big(v(n)+b_{m}\big)$
is either one or zero, and $b_{m}$ are fixed and uniformly distributed
between $-b_{\max}$ and $b_{\max}$, there are $N+1$ possible combinations
of zeros and ones in the $N$ inputs to $w_{m}$, and these combinations
contain consecutive zeros followed by consecutive ones (except for
two combinations with only zeros or only ones). For example, with
$N=4$, the only possible combinations are $0000$, $0001$, $0011$,
$0111$, and $1111$. This means that the $N$ multiplications by
$w_{m}$ followed by the corresponding $N-1$ additions, and the addition
of $c_{0}$, can be implemented with a look-up table of size $N+1$
and with entries $u_{q}$, $q=0,1,\ldots,N$, where
\begin{equation}
u_{q}=\begin{cases}
c_{0}, & q=0,\\
c_{0}+\sum_{i=N-q+1}^{N}w_{i}, & 1\leq q\leq N.
\end{cases}\label{eq:memory-entries}
\end{equation}

\begin{figure}
\begin{centering}
\includegraphics[scale=0.7]{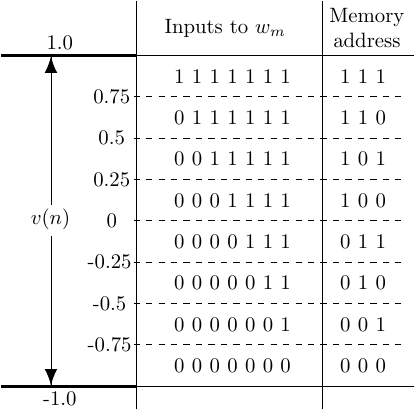}
\par\end{centering}
\caption{Signal levels and corresponding inputs to $w_m$ and memory address for $N=7$.}
\label{Flo:signal-mapping}
\end{figure}

\begin{figure}
\begin{centering}
\includegraphics[scale=0.55]{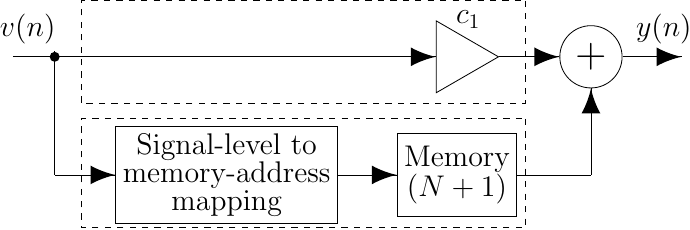}
\par\end{centering}
\caption{Implementation of the proposed linearizer using a memory of size $N+1$ and $\lceil \log_2(N+1) \rceil$ memory-address bits (see Footnote 2).}
\label{Flo:proposed-implementation}
\end{figure}

\begin{figure}[t]
\begin{centering}
\includegraphics[scale=0.55]{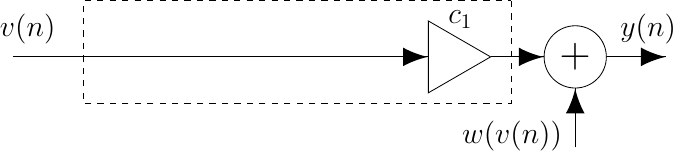}
\par\end{centering}
\caption{Equivalence of the implementation in Fig. \ref{Flo:proposed-implementation}.}
\label{Flo:proposed-implementation-equivalence}
\end{figure}

Secondly, with $b_{\max}$ as in \eqref{eq:b_max}, and with $b_{m}$
uniformly distributed between $-b_{\max}$ and $b_{\max}$, it is
readily shown that the whole signal-value region from $-1$ to $1$,
of the input signal $v(n)$, can be divided into $N+1$ consecutive
sub-regions of the same size $2/(N+1)$, with each sub-region corresponding
to a unique combination of the $N+1$ possible zero/one combinations
at the quantizers' outputs. This is exemplified in Fig. \ref{Flo:signal-mapping}
for $N=7$, and it means in general that neither the $N$ additions
by the bias values $b_{m}$ nor the $1$-bit quantizations and subsequent multiplications have to be
carried out. Instead, the look-up table can be directly
addressed by identifying and mapping the input signal level to a corresponding
memory address. For this, it suffices to use $\lceil \log_2(N+1) \rceil$ memory-address bits as exemplified in Fig. \ref{Flo:signal-mapping}. The proposed linearizer can therefore be implemented
as Fig. \ref{Flo:proposed-implementation} depicts\footnote{By scaling the linear and nonlinear branches by $1/c_1$, one could remove the multiplication in the linear branch. However, in a practical implementation, scaling of the signal levels is required \cite{Jackson_96}, which means that one multiplication is generally needed anyhow.}. It can be
equivalently described by the scheme in Fig. \ref{Flo:proposed-implementation-equivalence},
since the look-up table based implementation corresponds to adding
a time-varying value $w\big(v(n)\big)$ which can take on the $N+1$
values in the table and whose value for each $n$ is determined by
the value of $v(n)$. Before proceeding, it is noted that post-correction
methods for ADCs utilizing look-up tables have appeared earlier, as
in \cite{Handel_2000} for the memoryless case. However, those methods
are different as they target small-scale nonlinearities caused by
nonideal quantizers (deviations from uniform quantizations) and the
correction table is then designed based on ADC output-code statistics.
Further, those correction methods (with one fixed table) typically
offer modest signal-to-noise-and-distortion ratio (SNDR) improvements
for some frequencies but can even deteriorate the SNDR for other frequencies
\cite{Handel_2000}. The linearizer in this paper
targets more general nonlinearities and larger SNDR improvements,
and it is designed to correct all frequencies in a prespecified frequency
band. To further stress the difference, it is noted that 1) the proposed
linearizer is designed via the basic structure in Fig. \ref{Flo:proposed-scheme}
and it can also be implemented in that way (but the look-up table
implementation can be more efficient), which is not the case for the previous
look-up-table-based methods, and 2) the size of the memory ($N+1$)
in the proposal can be kept relatively small as it is determined by
the number of nonlinear branches ($N$), and not by the number of data-bits combinations utilized when designing the tables in the previous methods.

\subsection{Design \label{sec:Design}}
An advantage of the proposed linearizer, as well as that in \cite{deiro23},
over traditional neural networks is that it can be readily designed
through matrix inversion, thereby eliminating the costly and time-consuming numerical optimization that is traditionally used when training
neural networks. Further, an advantage of the proposed linearizer,
over that in \cite{deiro23}, is that it requires only one design
(one matrix inversion) as $b_{\text{max}}$ and thereby $b_{m}$ are
predetermined according to \eqref{eq:b_max} and \eqref{eq:b_m}.
In \cite{deiro23}, in contrast, a number of designs (matrix inversions)
were carried out, one for each value of $b_{\text{max}}$ over a set
of different values, and the best of the so obtained solutions was
selected.

Since the bias values $b_{m}$ are predetermined in the proposed linearizer,
the design amounts to determining the parameters $c_{0}$, $c_{1}$,
and $w_{m}$, $m=1,2,\ldots,N$, so that the output signal $y(n)$
approximates the desired signal $x(n)$ as well as possible, here
in the least-squares sense. Even if the proposed linearizer can be
implemented with a look-up table, its design is most easily carried
out via the basic scheme in Fig. \ref{Flo:proposed-scheme} with the
nonlinear functions in \eqref{eq:one-bit-quantizer} and bias values
in \eqref{eq:b_max} and \eqref{eq:b_m}. Further, in the design,
$c_{1}v(n$) in \eqref{eq:prop_linearizer} is replaced with $v(n)+\Delta c_{1}v(n)$
and then $\Delta c_{1}$ is determined. Thereby, all parameters to
be determined are small (zero when the distortion is zero). In addition,
$L_{2}$-regularization is used to reduce the risk of ill-conditioned
matrices and avoid large parameter values. Based on the above, the
following design procedure is proposed.
\begin{enumerate}
\item Generate a set of $R$ reference signals $x_{r}(n)$ and their distorted
signals $v_{r}(n)$, for $r=1,2,\ldots,R$, utilizing a signal model
such as that in \eqref{eq:signal-model} or measured data.
\item Minimize the cost function $E$ defined by
\begin{equation}
E=\frac{1}{RL}\sum_{r=1}^{R}\sum_{n=1}^{L}\big(y_{r}(n)-x_{r}(n)\big)^{2},\label{eq:E}
\end{equation}
where $L$ denotes the length of the signal. To solve this problem,
define ${\bf w}$ as a $(N+2)\times1$ column vector encompassing
all coefficients $w_{m}$ for $m=1,2,\ldots,N$, $c_{1}$, and $c_{0}$,
and let ${\bf{A}}_{r}$ be an $L\times(N+2)$ matrix with columns $l$
for $l=1,2,\ldots,N+2$, filled with the $L$ samples $f_{m}\big(v_{r}(n)+b_{m}\big)$
for $m=1,2,\ldots,N$, $L$ samples of $v_{r}(n)$, and $L$ ones
for the constant $c_{0}$. Minimizing $E$ in \eqref{eq:E} in the
least-squares sense then gives the solution
\begin{equation}
{\bf w}={\bf A}^{-1}{\bf b},
\end{equation}
where
\begin{equation}
{\bf A}=\lambda{\bf I}+\frac{1}{RL}\sum_{r=1}^{R}{{\bf{A}}_{r}^{\top}{\bf A}_{r}},\quad{\bf b}=\frac{1}{RL}\sum_{r=1}^{R}{{\bf A}_{r}^{\top}{\bf b}_{r}}.
\end{equation}
Here, ${\bf{A}}_{r}^{\top}$ is the transpose of ${\bf{A}}_{r}$, ${\bf {b}}_{r}$
is an $L\times1$ column vector with the samples $x_{r}(n)-v_{r}(n)$,
and $\lambda{\bf I}$ is a diagonal matrix with small diagonal values
$\lambda$ for the $L_{2}$-regularization. The linearized output (row vector)
${\bf y}_r=y_{r}(n)$, $n=1,2,\ldots,L$, is given by
\begin{equation}
	{\bf{y}}_r={\bf{v}}_r+{\bf{w}}^{\top}{\bf{A}}_{r}^{\top}.
\end{equation}
where ${\bf v}_r=v_{r}(n)$, $n=1,2,\ldots,L$ is also a row vector.
\item Assess the performance of the linearizer over $M$ signals, where
$M\gg R$, to ensure a robust validation. 
\end{enumerate}

\section{Evaluations and Comparisons\label{sec:Evaluation-and-Results}}
For the evaluations and comparisons, we assume the same distorted
signal $v(n)$ as in \cite{deiro23} where $v(n)$ is modeled as in
\eqref{eq:signal-model} with $a_{0}=0$, $a_{1}=1$, and $a_{p}=(-1)^{p}\times0.15/p$,
for $p=2,3,\ldots,P$ with $P=10$. 

\begin{figure}
\centering \includegraphics[scale=0.84]{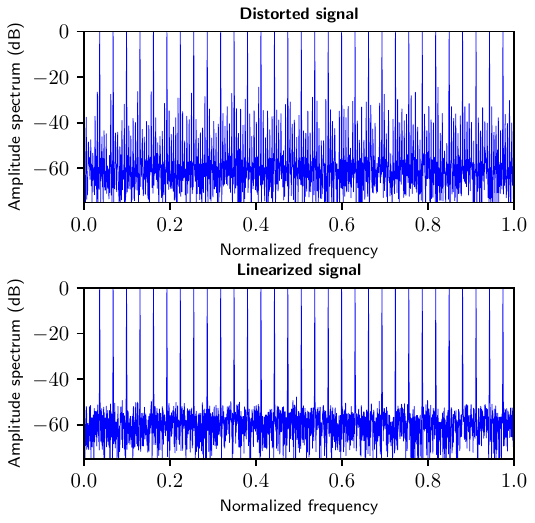}
\caption{Spectrum before and after linearization for a multi-sine signal using
the proposed linearizer with $N=32$ nonlinear branches.}
\label{Flo:Spectrum_OFDM}
\end{figure}

\begin{figure}
\centering \includegraphics[scale=0.83]{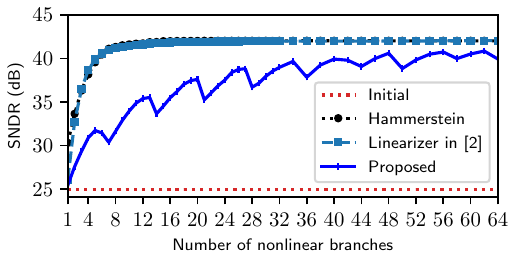}
 \caption{SNDR versus the number of nonlinear branches.}
\label{Flo:SNDR_versus_N}
\end{figure}

\textit{Example 1: }We consider the multi-tone signal
\begin{equation}
x(n)=G\times\sum_{k=1}^{31}A_{k}\sin(\omega_{k}n+\alpha_{k}),
\end{equation}
where $A_{k}=1$ for all $k$, and $\alpha_{k}$ are randomly chosen
from $\{\pi/4,-\pi/4,3\pi/4,-3\pi/4\}$, which corresponds to QPSK
modulation. The frequencies $\omega_{k}$ are given by
\begin{equation}
\omega_{k}=\frac{2\pi k}{64}+\Delta\omega,
\end{equation}
in which case the signal corresponds to the quadrature (imaginary)
part of $31$ active subcarriers in a $64$-subcarrier OFDM signal with a random
frequency offset $\Delta\omega$. In the design and evaluation we
use, respectively, $R=1$ and $M=2500$ signals with randomly generated
frequency offsets assuming uniform distribution between $-\pi/64$
and $\pi/64$, quantized to $8$ bits, and of length $L=8192$. The
gain $G$ is selected so that the distorted signal is below
one in magnitude. For the $L_{2}$-regularization, we use $\lambda=0.0002$.
Figure \ref{Flo:Spectrum_OFDM} plots the spectrum before and after
linearization for one of the signals. Figure \ref{Flo:SNDR_versus_N}
plots the mean SNDR over 2500 signals for each linearizer instance (with an SNDR variance of 
some 0.5 dB for all instances) versus the number
of branches for the proposed linearizer, the one in \cite{deiro23},
and the Hammerstein linearizer, designed in the same least-squares sense (for all designs, the optimized parameter
values were quantized to $12$ bits). For the signals considered in
this example, quantized to $8$ bits, the signal-to-noise ratio (SNR)
is some $42$ dB without distortion, and the SNDR is some $25$ dB
for the distorted signal before the linearization. Hence, the linearizer
can at most improve the SNDR by $17$ dB in this example, which corresponds
to almost $3$ bits improvement.

\begin{figure}
\centering \includegraphics[scale=0.82]{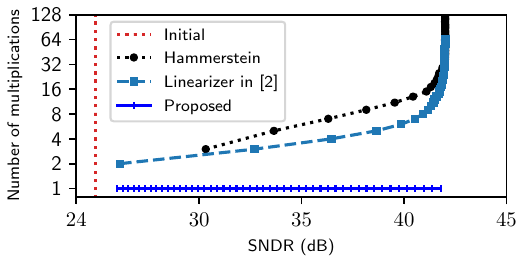}
\caption{SNDR versus the number of multiplications required per corrected output
sample.}
\vspace{-5px}
\label{Flo:SNDR_versus_mult}
\end{figure}

As seen in Fig. \ref{Flo:SNDR_versus_N}, for the proposed linearizer,
the SNDR does not increase monotonically. This is because its bias
values are not optimized but instead selected according to \eqref{eq:b_max}
and \eqref{eq:b_m} to enable the look-up table implementation. Further,
the proposed linearizer requires more nonlinear branches than the
other two methods to reach the same SNDR. However, to assess the implementation
complexity, one should not consider the number of branches but instead
the complexity required to linearize each output sample. With $N$
nonlinear branches, the Hammerstein linearizer (where $N=K-1)$ requires
$2N+1$ multiplications and $N+1$ additions whereas the linearizer
in \cite{deiro23} requires $N+1$ multiplications and $2N+1$ additions.
In general, multiplications are considerably more complex to implement
than additions and the linearizer in \cite{deiro23} is thus more
efficient than the Hammerstein linearizer for the same number of nonlinear
branches. The proposed linearizer, implemented with a look-up table,
requires only one multiplication and one addition per output sample.
Hence, in terms of computational complexity, this linearizer clearly
outperforms the other two, as illustrated in Fig. \ref{Flo:SNDR_versus_mult}
which plots the number of multiplications versus the SNDR for the
three methods. However, the proposed linearizer also have the additional
cost of the memory look-up implementation, which is 
dependent upon the hardware platform.

\begin{figure}
\centering \includegraphics[scale=0.82]{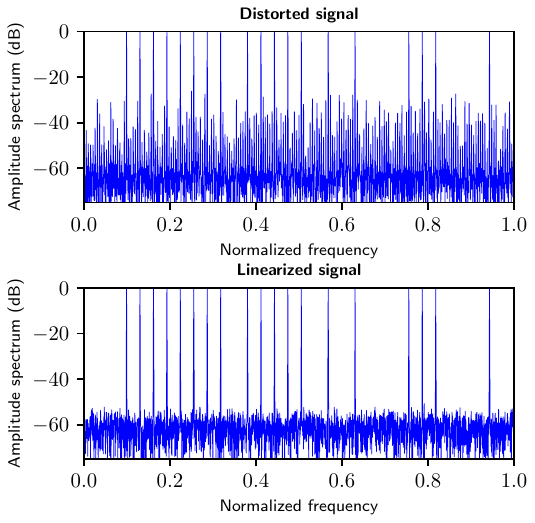}
\caption{Spectrum before and after linearization for a multi-sine signal with
null subcarriers using the proposed linearizer with $N=32$ nonlinear
branches.}
\vspace{-5px}
\label{Flo:Spectrum_OFDM_zeros}
\vspace{-5px}
\end{figure}

\begin{figure}[t]
\centering \includegraphics[scale=0.82]{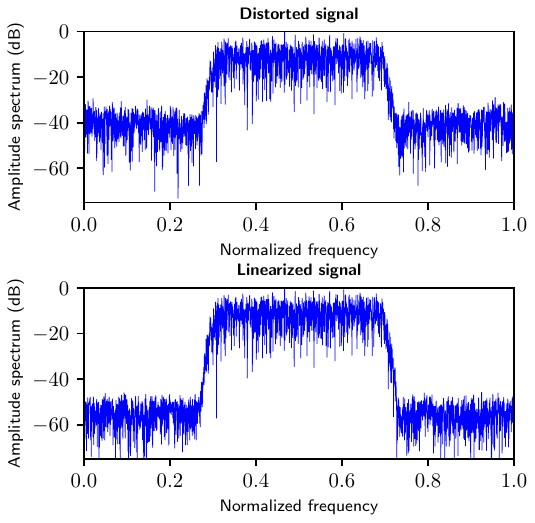}\caption{Spectrum before and after linearization for a bandpass filtered white-noise
signal using the proposed linearizer with $N=32$ nonlinear branches.}
\vspace{-5px}
\vspace{-5px}
\label{Flo:Spectrum_FRN}
\end{figure}

\textit{Example 2: }To further illustrate the robustness of the proposed
linearizer designed in Example 1, we have also evaluated it for the
same type of multi-sine signal as in Example 1 but with some of the
subcarriers set to zero, and a bandpass filtered white-noise signal covering
50\% of the Nyquist band. As illustrated in Figs. \ref{Flo:Spectrum_OFDM_zeros}
and \ref{Flo:Spectrum_FRN} for one of each of these signals, essentially
the same result is obtained. Less than 1 dB SNDR degradation compared
to the linearized signals considered in Example 1 was observed.

\section{Conclusions\label{sec:Conclusions}}
This paper introduced a low-complexity memoryless linearizer for suppression
of distortion in analog frontends. It is based on our recently introduced
linearizer in \cite{deiro23}, which is inspired by neural networks
but offers orders-of-magnitude lower complexity, and can outperform
the conventional parallel memoryless Hammerstein linearizer. Further,
it can be designed through matrix inversion and thereby the costly
and time consuming numerical optimization traditionally used when
training neural networks is avoided. The proposed linearizer is different
from \cite{deiro23} in that it uses 1-bit quantizations as nonlinear
activation functions (instead of ReLU and modulus operations) and
different and carefully selected bias values. These features enable
a look-up table implementation (Fig. \ref{Flo:proposed-implementation})
which means that the linearization only requires one multiplication
and one addition to correct each output sample, regardless of the
number of nonlinear branches ($N$). 

Examples included also demonstrated that the same SNDR can be reached
with the proposed linearizer, as with the previous linearizers, by
increasing the number of nonlinear branches. For the same SNDR improvement,
the proposed linearizer is thus superior in terms of computational
complexity, but it also requires the additional memory. Hence, there is a trade-off between computational
complexity and the additional cost of the memory look-up implementation. Again,
it is stressed though that the size of the memory ($N+1$) in the
proposal is relatively small since it is determined by the number
of nonlinear branches ($N$), not by the number of data-bits combinations
utilized when designing the tables in the previous look-up-table-based
methods. Further studies thus include the utilization of both $1$-bit
quantization and ReLU/modulus operations in the linearizer in order
to find the lowest implementation complexity, given that the cost
of different operations is available for a specific hardware platform.
The extension to memory linearizers addressing frequency dependent
nonlinearities will also be studied.

\section*{Acknowledgment}
This work belongs to the project Baseband Processing for Beyond 5G Wireless (B02), financially supported by ELLIIT.
\vspace{-2px}

 \bibliographystyle{IEEEtran}
\bibliography{references/bibliography}

\end{document}